\documentclass[prl,twocolumn]{revtex4}%
\usepackage{eurosym}
\usepackage{graphicx}
\usepackage{dcolumn}
\usepackage{ulem}
\usepackage{textcomp}
\usepackage{amssymb}
\usepackage{amsmath}
\usepackage{color}
\usepackage{footnote}
\usepackage{microtype}
\usepackage{amsfonts}%

\providecommand{\U}[1]{\protect\rule{.1in}{.1in}}
\newcommand{\beq}{\begin{equation}}
\newcommand{\eeq}{\end{equation}}
\newcommand{\beqa}{\begin{eqnarray}}
\newcommand{\eeqa}{\end{eqnarray}}

\begin{document}
\title{Spin Spiral and Topological Hall Effect in a Metamagnet Fe$_{3}$Ga$_{4}$}
\author{Mahdi Afshar, and Igor I. Mazin}
\affiliation{Department of Physics and Astronomy, and Quantum Science and Engineering
Center, George Mason University, Fairfax, VA 22030 }

\begin{abstract}
A new mechanism for Topological Hall Effect (THE) was recently proposed for
the spiral magnet YMn$_{6}$Sn$_{6},$ which requires transverse conical spiral
(TCS) magnetism, induced by external magnetic field, combined with thermally
excite helical spiral magnons. In principle, this mechanism should be
applicable to other itinerant spiral magnets as well. In this paper, we show
that another metamagnetic compound, Fe$_{3}$Ga$_{4}$, where THE was observed
experimentally before, in one of its phases satisfies this condition, and the
proposed theory of thermal-fluctuation driven THE is quantitatively consistent
with the experiment. This finding suggests that this mechanism is indeed
rather universal, and the effect may have been observed in other compounds
before, but overlooked.

\end{abstract}
\maketitle

\section{I. Introduction}

During the last decades, topological effects driven by magnetic textures have
attracted considerable attraction \cite{Zang2019, Nagaosa2013, Nagaosa2012,
Bradlyn2016, Armitage2018}. In particular, the Hall effect has been widely
used as a probe for topological effects. In the classical Hall effect,
discovered more than a century ago, the Lorentz force resulting from an external
magnetic field gives rise to
an electric field perpendicular to the electron current. The theory of this
phenomenon is well known, and stipulates that the effect is linear in the
magnetic field, with the ordinary Hall resistivity $\rho^{O}=R_{0}H$ (the
proportionality coefficient $R_{0}$ depends on the details of the Fermi
surface). In systems with broken time-reversal symmetry (for instance, in
ferromagnetic), there exists another contribution to the off-diagonal
resistivity, dubbed \textquotedblleft anomalous Hall effect\textquotedblright%
\ (AHE), $\rho^{A}=R_{s}M$. This contribution is proportional to the
magnetization $M$, and gives rise to a Hall effect even in the absence of an
externally applied magnetic field. While this relation is not always
true, for instance, it is violated in some antiferromagnets \cite{Libor}, it
has been routinely used to identify the AHE in the experiment.

Very recently an additional mechanism generating an off-diagonal resistivity
in magnets with noncoplanar moments was identified \cite{Nagaosa}. Interestingly,
contrary to the AHE, this mechanism does not require spin-orbit interaction,
albeit can benefit from it \cite{Batista}. This mechanism, often called
Topological Hall Effect (THE) is based on the Berry phase an electron acquires
when its spin follows a spatially varying magnetization that is present in
such materials. It was shown that its amplitude is proportional to the
so-called scalar spin chirality (SSC), defined as the triple product of three
spins forming a triangle:%
\begin{equation}
\Omega=\mathbf{S}_{1}\cdot(\mathbf{S}_{2}\times\mathbf{S}_{3})
\end{equation}

In principle, this mechanism is not supposed to work in system with zero SSC,
 and weak spin-orbit (as in many 3d metals). Yet, in several cases
sizeable deviations from the standard formula, $\rho=\rho^{O}+\rho^{A}%
=R_{o}H+R_{s}M,$ were reported \cite{ghimire,mendez,LMG,wang},
and ascribed to THE, even though for all these system the magnetic structure is known and does
not have any SSC.

For one of this compound, namely YMn$_{6}$Sn$_{6},$ particularly detailed set
of experimental data was available \cite{ghimire}, and another mechanism for
THE was proposed. Within this scenario, SSC emerges through a fluctuational
mechanism akin to the emerging nematicity in Fe-based superconductor \cite{Jorg}. The
resulting THE amplitude grows roughly linearly with temperature, with a
quadratic dependence on magnetization. The prerequisites to this fluctuational
THE (fTHE) are (a) a transverse conical spiral magnetic state at least in some
range of temperatures and external fields (b) itinerant electrons strongly
coupled with this spiral (ideally, formed by the same electron orbitals) and
(c) strong fluctuations.

In this paper, we will study another compound where THE has been
reported \cite{mendez}, Fe$_{3}$Ga$_{4}$, and will show that this observation is consistent with the
same fTHE mechanism. In the following section we will describe the
compound and the experimental picture, then we will present the results of our Density Functional Theory (DFT)
calculations and discuss  the magnetic phase diagram. After that, we
will review the theory of the fTHE and apply it to Fe$_{3}$Ga$_{4}.$

\section{II. Experimental situation}

\begin{figure}[tbh]
\begin{center}
\includegraphics[width=0.85 \columnwidth]{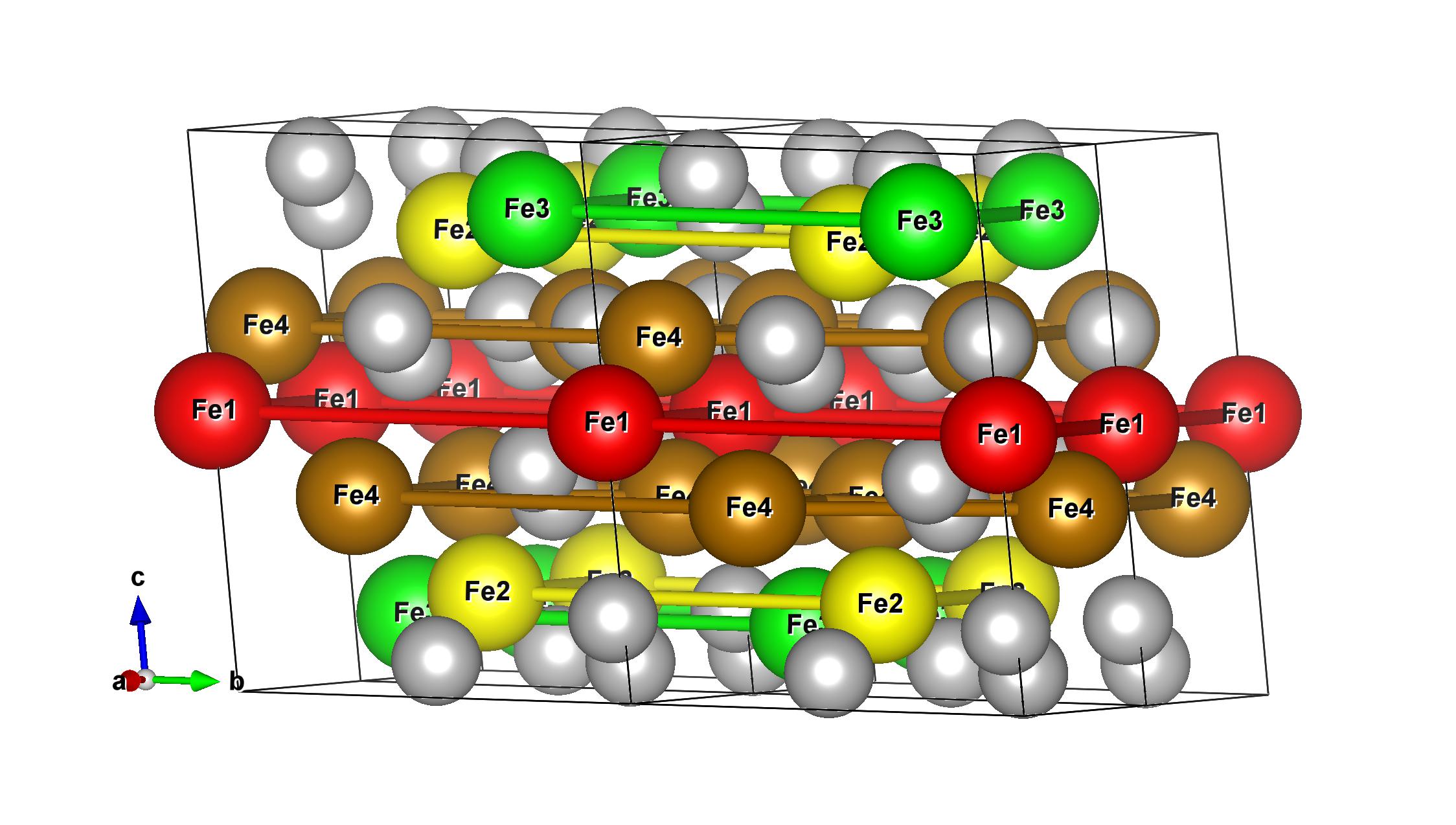}
\end{center}
\caption{(Color online) Layered structure of Fe atoms in Fe$_{3}$Ga$_{4}$
crystal structure. The four different crystallographically inequivalent iron
sites are shown in different colors. There are also four unique Ga sites, which
are all shown in grey.}%
\label{str}%
\end{figure}

Fe$_{3}$Ga$_{4}$ crystallizes in a base-centered monoclinic structure, with
the symmetry group C2/m, and a rather complex primitive unit cell of three
formula units. The four crystallographically inequivalent Fe sites form seven
parallel sheets along the c-direction as shown in Fig. \ref{str}, with
interlayer distances of 0.368, 1.334, 1.104, 1.104,1.334, 0.368 and 0.977
\AA . The lattice parameters are $a=10.0979$ \AA , $b=7.6670$ \AA , and
$c=7.873$3 \AA ~ with an obtuse angle of $\beta=106.298^{\circ}$
\cite{mendez}. While crystallographically and electronically, as will be
discussed in more details later, it is rather 3D, magnetically it can be
viewed as a stack of ferromagnetically ordered planes with complex, but,
presumably, weaker interplanar coupling \cite{Wu}.

The material is known to have two metamagnetic transitions \cite{mendez,Wu},
from a ferromagnetic (FM) to a spin density wave (SDW) at $T_{1}=60$ K, and
back to a ferromagnetic state at $T_{2}\approx360$ K (in this paper we apply
the term SDW to any phase where spin polarization varies periodically in space;
thus defined SDW can be either a spiral, or an \textit{amplitude }SDW, wherein
the magnitude of the magnetic moment varies continuously, or a combination of
both). The long-range order is lost at $T_{3}=420$ K. The nature of the SDW
phase will be discussed later, we will just mentioned that the neutron data can
be fit equally well \cite{Wu, cao} by an \textit{amplitude }SDW, where the spins
are mostly aligned along $c,$ or by a spin spiral, with the helical orientation,
$i.e.,$ with the spins rotating in the $ab$ plane. Either way, the spiral wave
vector appears to be $(0,0,0.29).$ The low-$T$ and 
the high-$T$ phases that are
identified as ferromagnetic should be more correctly characterized as
uncompensated magnetic phases. Especially the high-$T$ phase may be a
noncollinear canted phase with zero net magnetization. In this paper, however,
we will not be concerned with the natures of those phases, but only with the
SDW phase between $T_{1}$ and $T_{2}.$

Experimentally, the low-temperature ferromagnetic phase shows the lowest
susceptibility in the fields below 0.3 T for the field direction along $a,$
and the highest along $c,$ but the $c$ and $b$ directions are nearly the same.
This was interpreted \cite{Wu, cao} as if $a$ is the easy axis, and $b$ is the
hard one. On the first glance, this implies that the ground state in the SDW
phase can be either amplitude wave with polarization along $a$, or a cycloidal
spiral with the spins rotating in the $ac$ plane. However, the latter is not
compatible with the neutron data, therefore the authors of Ref. \cite{Wu, cao}
argued it must be an amplitude wave (note that in text of Ref. \cite{Wu} it
was incorrectly stated that the suggested SDW is linearly polarized along $c,$
but the figure correctly shows the one polarized along $a$ \cite{cao}). However,
the difference between the $c-$ and $b-$axis susceptibilities being at least
five times smaller than that between $c$ and $a,$ an $ab$ spin helix could not
be reliably excluded.

While not explicitly discussed in Ref. \cite{mendez}, magnetometry clearly
shows a spin-flop transition for the field $H\perp c,$ which is not very
distinctly defined (probably because of the sample quality), but is seen to occur
in the SDW phase at the field $H_{b}\gtrsim1$ T at $T=150$ K, with $H_{b}$
linearly decreasing with temperature down to $\sim0.5$ T at $T\sim300$ K.
Careful examination of the $M(H)$ curves for $H\perp c$ suggests a possibility
of another spin-flop transition, at very low fields $H_{a}\lesssim0.1$ T, but
that is rather speculative.

The residual resistivity was relatively large, with the room-temperature ratio
$\sim2,$ indicating a large number of defects and possibly deviations from
stoichiometry. The residual specific heat coefficient $C(T)/T|_{T\rightarrow
0}=23$ mJ/mole K$^{2},$ corresponding to the density of states (DOS) at the
Fermi level $N(0)\approx10$ states/f.u.. Only the first phase transition, at
$T=T_{1},$ has a distinct specific heat signature, and the entropy change is
very small, less than 0.3\% of $R\log2$, indicating that the transition
occures between two well ordered states. The authors of Ref. \cite{mendez}
estimate that entropy change between $T_{2}$ and $T_{3}$ as 0.43 J/mole K,
which is less than 10\% of $R\log2,$ consistent with a quasi-2D character of
magnetism in this material.

Transport measurements indicate an extra contribution for the Hall effect
$\rho_{xy}$ (\textit{i.e., }in a magnetic field in the $ab$ plane) for an
intermediate temperature range, roughly coinciding with the $(T_{1},T_{2})$
interval, compared with the standard combination of an anomalous and an
ordinary Hall effect,
\begin{equation}
\rho_{xy}(H)=R_{o}H+R_s M.
\end{equation}
The coefficients $R_{o}$ and $R_{s}$ strongly depend on the phase, and, inside
each phase, also depend on temperature, which makes it difficult to quantify
the additional, presumably topological, contribution, but one can say with
confidence that this contribution increases with temperature up to the highest
reported temperature of 350 K.

\section{III. DFT Calculations}

Calculations of the structural, electronic, and magnetic properties of bulk
Fe$_{3}$Ga$_{4}$ were performed using the Vienna {\it ab initio} simulation package
(VASP) \cite{kresse93,kresse94,kresse961,kresse962}. Fe 3s, 3p, 3d, and 4s and Ga 3p, 3d, and 4s states were
treated as valence. The plane cut-off was 500 eV. We use the Gaussian smearing
with the width of 0.05 eV, this value ensuring an entropy contribution to the free
energy of less than 1 meV/atom. The generalized gradient approximation (GGA) was
used for the exchange-correlation functional \cite{gga}. The spin-orbit coupling
(SOC) was included in the self-consistent calculations, unless mentioned
otherwise. The k-point sampling was based on a $\Gamma$-centered grid for all
calculations and we used an optimized (10$\times$10$\times$10) k-points, except
when for the density of states (DOS) calculations, where the 12$\times$12$\times$9 grid was utilized.

In addition, we used an all-electron Full-Potential Local Orbitals (FPLO) \cite{fplo}
package. which solves the fully relativistic Dirac equations \cite{ksd}. The
basis set included Fe(1s, 2s, 2p, 3s, 3p, 3d), and Ga(3s, 3p, 3d, 4s, 4p, 4d,
5s) states. The total energy converged to 0.001 meV. In order to
address the possible effect of the on-site electron correlations, we employed
the GGA+U method in the fully-localized limit \cite{pickett}. As implemented in FPLO, it has
full nonspherical double counting subtraction (as opposed to most other codes),
whereby the first Slater integral is defined as $F_{0}=U$, where $U$ is the
Hubbard repulsion, and the Hund's rule coupling defined the other integrals
via $J=(F_{2}+F_{4})/14$, and the ratio of $F_{4}/F_{2}$ is set to 0.625,
typical for 3d transition metals \cite{anisimov}. We used $J=0.9$ eV, and
varied $U.$

Spin spiral and unrestricted noncollinear calculations were performed using
the VASP package. For the former, the generalized Bloch theorem formalism
\cite{sandratski} was utilized, and verified against 1$\times$1$\times$4
unrestricted noncollinear calculations. By construction, the spiral formalism
does not include the spin-orbit coupling, but relevant energy differences were
similar to those in relativistic supercell calculations.

\begin{figure}[tbh]
\includegraphics[width=0.95 \columnwidth]{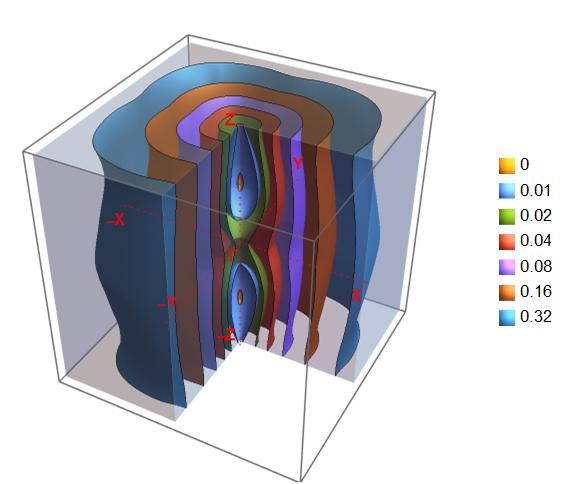}
\includegraphics[width=0.95 \columnwidth]{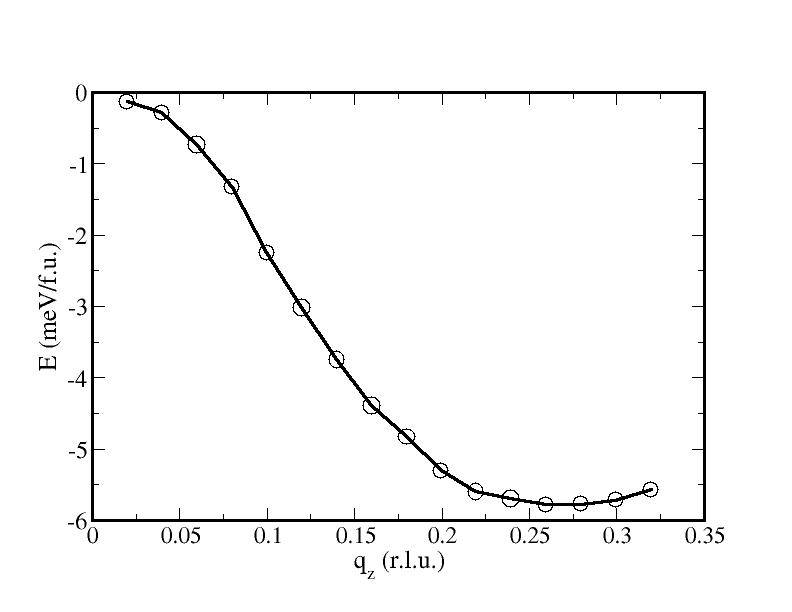}
\caption{(Top panel) 3D contour plot of the total energy of a non-relativistic spiral with a spiral vector
$\mathbf{q}=(x,y,z)$, where $x,y,z$ are components in reciprocal lattice
coordinates. (Bottom panel) same, for the vector $\mathbf{q}=(0,0,z)$.}%
\label{spiral}%
\end{figure}

Fig. \ref{spiral} summarizes the result of these calculations. We have scanned
the irreducible part of the primitive Brillouin zone using the 
5$\times$5$\times$4 mesh with the step of 0.1 $G$ from 0 to $0.5G$ for each
crystallographic direction ($G$'s are the corresponding reciprocal lattice
vectors), altogether 216 calculations. One ca see that the magnon spectrum is stiff
along $x$ and $y,$ and soft along $z,$ with a minimum close to $\mathbf{q}%
=(0,0,0.27)$ in reciprocal lattice units. We then calculated the spiral
energies with a finer mesh of 7$\times7\times7,$ along the line $\mathbf{q}%
=(0,0,q_{z}),$ with a step of 0.02 in $q_{z}$ (Fig. \ref{spiral}).The position
of the minimum has not change. The value of $\mathbf{q}=(0,0,0.27)$ agrees
well with the experimental number.

\begin{figure}[tbh]
\begin{center}
\includegraphics[width=0.98 \columnwidth]{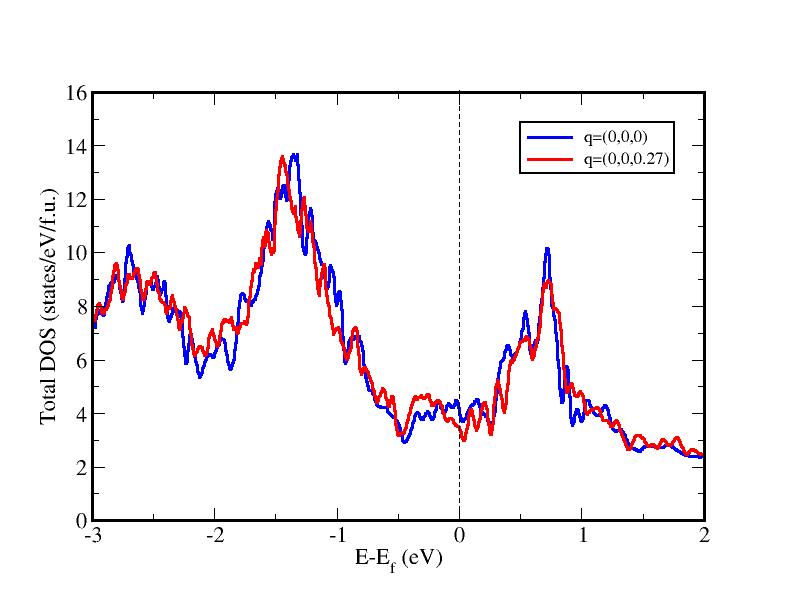}
\end{center}
\caption{Density of states near the Fermi level for the ferromagnetic
[$\mathbf{q}=(0,0,0)]$ and spiral [$\mathbf{q}=(0,0,0.27)$] states. Note
small, but discernible weight transfer away from the Fermi level.}%
\label{DOS}%
\end{figure}

We have also tried to stabilize an amplitudinal SDW, as suggested in Ref.
\cite{Wu}. It never stabilizes, indicating that the DFT ground state is
resoundingly spiral.

While the FM Fermi surface does not show any visible nesting feature, nor does
the non-interacting susceptibility (either $\chi_{zz}$ or $\chi_{+-})$ show
any well-defined maximum, the calculated density of states for the FM
($\mathbf{q}=0)$ and the spiral $\mathbf{q}=(0,0,0.27)$ states (Fig.
\ref{DOS}) show small spectral weight transfer from the region within a few
tenth of an eV near the Fermi level to farther energies, that is, a small, but
noticeable pseudogap effect.

\begin{figure}[tbh]
\begin{center}
\includegraphics[width=0.99 \columnwidth]{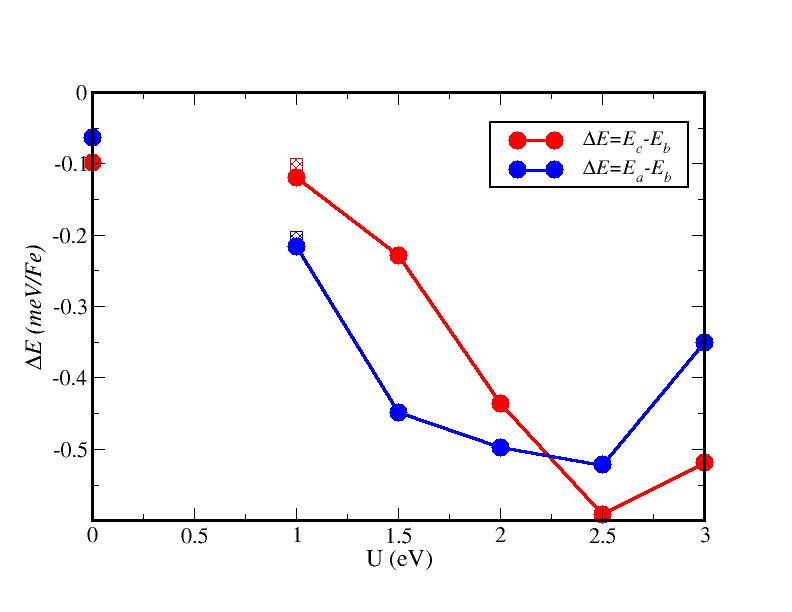}
\end{center}
\caption{Magneto-anisotropy energy for the quantization axis along the
crystallographic $a$, $b$ and $c$ axes. Calculations were performed in FPLO
for the Hund's of $J=0.9$ eV as a function of Hubbard $U$. Zero corresponds to DFT
calculations without the GGA+U correction. The two points for $U=1$ eV
correspond to the k-point meshes of $8\times8\times8$ and $12\times12\times
12$.}%
\label{mae}%
\end{figure}

We have also calculated the magnetic anisotropy as a function of the Hubbard
correction $U$ (calculations reported above were not including $U$). To this end, we used the FPLO method, which treats the
relativistic effects more accurately and the angular dependence of the GGA+U
term is included in a more systematic way. The results are presented in Fig.
\ref{mae}. The calculated anisotropy for the Hubbard parameter of $U$ from 1eV to 3eV and Hund's
$J=0.9$ eV agrees with the experiment. Also, particularly at $U\approx J,$ the
difference between the $c$ and $b$ orientations is on the order of 0.1 meV/Fe,
resulting in a stabilization energy for the cycloidal $ac$ spiral about 0.05
meV. 

One reason why the helical spiral may be more stable in the experiment
is the dipole-dipole interaction \cite{MM}. Indeed, in the long-wave-length
limit it contributes, for a cycloidal (but not helical) spiral an additional
energy equal to $\int\pi m^{2}dV,$ where $m$ is the magnetization density, and
the integration is over the entire crystal. Using the Fe$_{3}$Ga$_{4}$
parameters, we get an estimate of $0.06$ meV/Fe, comparable with, and slightly
larger than the electronic anisotropy energy.

\begin{figure}[tbh]
\includegraphics[width=0.99 \columnwidth]{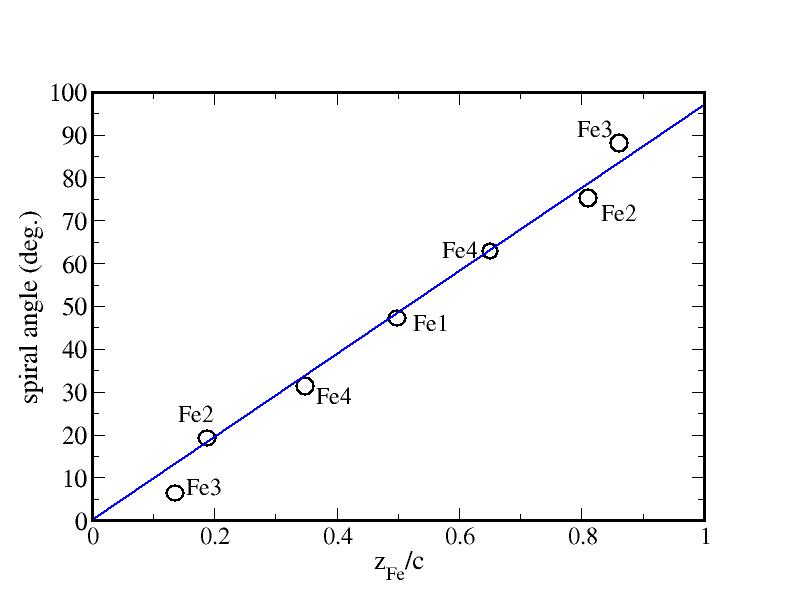} \caption{Spiral angle as
function of the position of an Fe layer within the unit cell, for the spiral
calculations with $\mathbf{q}=(0,0,0.27)$. No restrictions are imposed on the
magnetic moment directions within a single unit cell, while the consecutive
unit cells are rotated by 0.27$\times360^{\circ}$. The line shows the ideal
sinusoid, $\alpha= 0.27\times360^{\circ}z$. }%
\label{sinus}%
\end{figure}

In principle, the next step would be to attempt deriving a first principles
Heisenberg Hamiltonian. In Fe$_{3}$Ga$_{4},$ unfortunately, it is virtually
impossible because of too many inequivalent bonds and the fact that many
ferrimagnetic configurations simply fail to converge. On the other hand, it
appears that the SDW in Fe$_{3}$Ga$_{4}$ can be quite well described in a
continuous model. Indeed, as discussed above, a unit cell includes 9 Fe atoms
arranged in 7 separate $ab$ Fe layers stacked along $c.$ Our spin-spiral
calculations place no restriction on the mutual orientation of their magnetic
moments. Yet, the self consistent solution can be very accurately described by
a simple sinusoid, where the helix angle is given by $\alpha(z)=\cos
(0.27\times2\pi z/c)$ (Fig. \ref{sinus}). Only the two Fe3 layers slightly
deviate from this formula.

Interestingly, the calculated energy as a function of the spiral vector is
very well described by the function%
\begin{equation}
E=E_{0}+J_{1}\cos2\pi qh+J_{2}\cos4\pi qh,
\end{equation}
where $h=1.75,$ $J_{1}=3$ meV and $J_{2}=0.4$ meV, as if the Hamiltonian was
comprised of two antiferromagnetic Heisenberg interaction, one acting across
the distance of $1.75c$ and the other of $3.5c.$ Of course, in reality this
would be only an effective Hamiltonian, resulting from concerted action of all
sorts of exchange interactions, but it indicates that the overall magnetic
coupling is extremely long range.

In any event, the calculations unambiguously indicate that of the two possible
ground states compatible with the neutron scattering data it is the helical
spiral that is realized, and not an amplitude SDW.

\begin{figure*}[tbh]
\includegraphics[width=0.99\columnwidth]{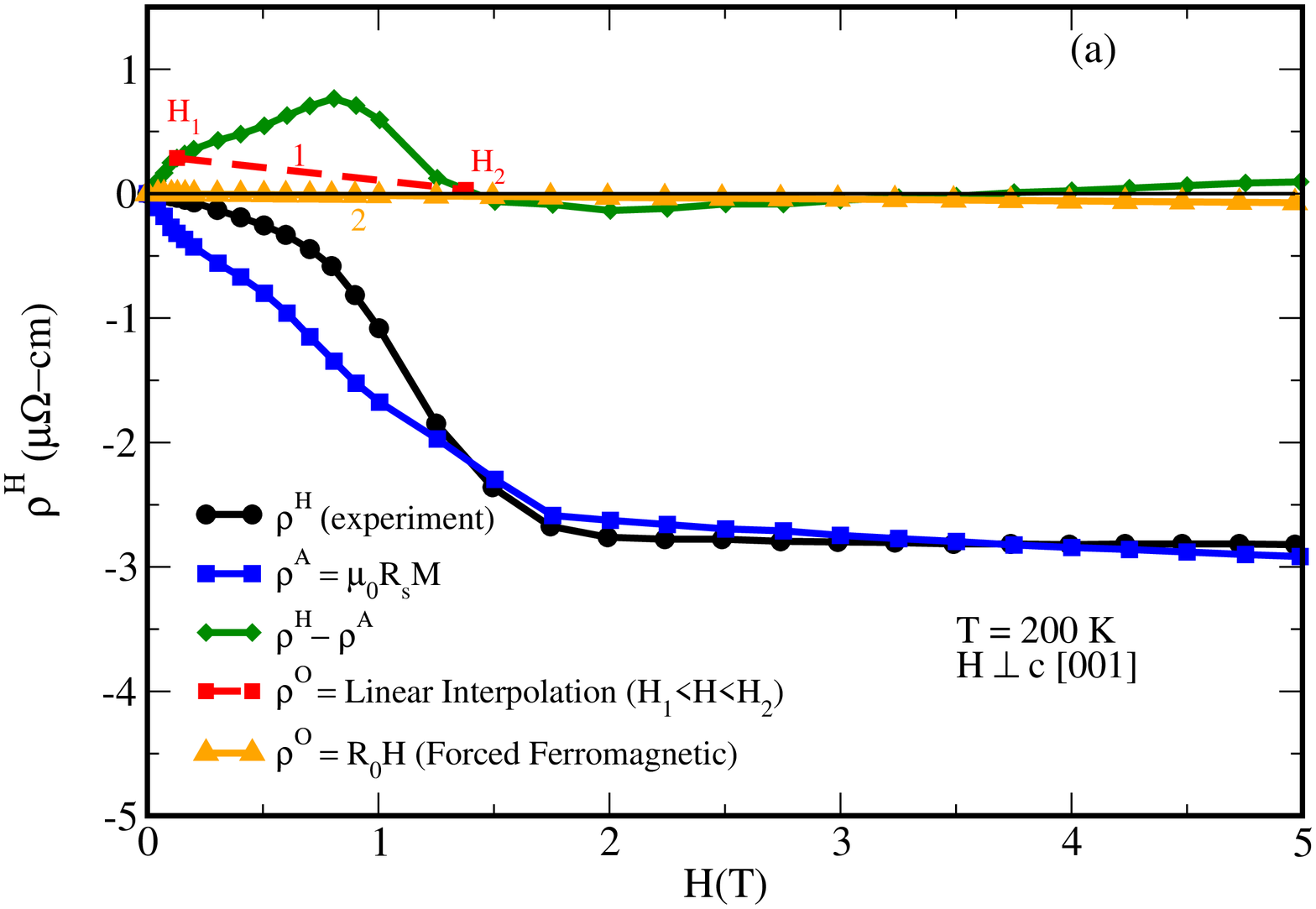}
\includegraphics[width=0.9182\columnwidth]{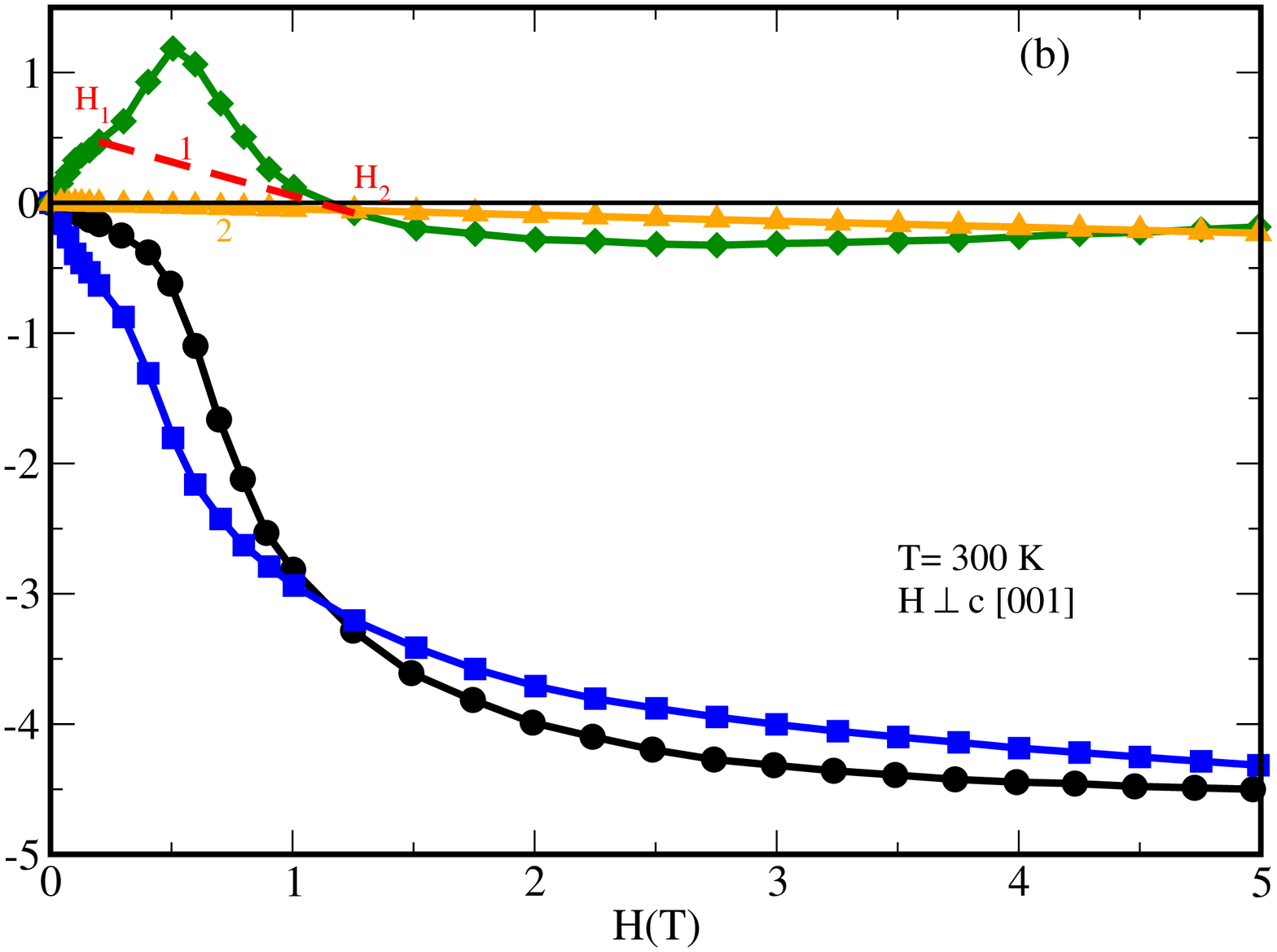} \caption{Suggested
decomposition of the Hall resisitivity measured by Mendez \textit{et al.}
\cite{mendez}, for two different temperatures.}%
\label{roab}%
\end{figure*}

\section{IV. Topological Hall Effect}

Typically, the Hall effect in metals is described as a sum of two components:
the ordinary Hall effect \cite{Hurd}, stemming from the Lorentz force
experienced by the charge carriers, and the anomalous Hall effect \cite{Hurd},
resulting from the interplay between the exchange field and spin-orbit
coupling. While there are notable exceptions (in particular, the anomalous Hall
effect was shown to exist even in some systems with zero
magnetization \cite{Libor}), it is customary to assume that the ordinary Hall
resistance is proportional to the applied field, $\rho^{O}=R_{0}H$, and the
anomalous to the net magnetization, $\rho^{A}=R_{s}M.$ Recently it was pointed
out that in noncoplanar magnets a third term should be added (see, for
instance, Ref. \cite{Nagaosa}, called topological Hall effect (THE),
proportional to the so-called scalar spin chirality $s,$ which can be defined
in a discrete representation as a triangular loop over near-neighbor magnetic
moment, $s=\mathbf{M}_{1}\cdot(\mathbf{M}_{2}\times\mathbf{M}_{3}).$

In the the continuous representation one can define the topological field,%
\begin{align}
b_{i}\mathbf{(r)}  &  =\sum_{jk}e_{ijk}\mathbf{M(r)\cdot}\left(
\frac{\partial\mathbf{M(r)}}{\partial r_{i}}\times\frac{\partial\mathbf{M(r)}%
}{\partial r_{k}}\right) \\
&  =\sum_{jk}\sum_{\alpha\beta\gamma}e_{ijk}e_{\alpha\beta\gamma}M_{\alpha
}\frac{\partial M_{\beta}}{\partial r_{i}}\frac{\partial M_{\gamma}}{\partial
r_{k}}%
\end{align}
where $i,j,k$ are Cartesian indices in the real space, and $\alpha
,\beta,\gamma$ in the spin space. This field can couple with the external
magnetic field and generate an additional contribution to the Hall resistivity
in the field parallel to $\mathbf{b}$ \cite{Nagaosa}. As a result, the Hall
resistivity is commonly written as:
\begin{equation}
{\rho}^{H}=R_{0}H+R_{s}M+{\rho}^{T}, \label{RH}%
\end{equation}

It is well known that a nonzero topological field $\mathbf{b}$ can be
generated by a linear combination of three (but not two) helical
spirals \cite{3}. It was recently pointed out\cite{ghimire} that a combination
of two spirals, one of which is helical, and the other transverse conical, can
have a nonzero topological field. Further more, Ghimire $et$ $al$\cite{ghimire} argued that
even if the ground state is a \textit{single} helical spiral propagating along
a given direction, say, $z,$ in a suitable magnetic field $\mathbf{H||x\perp
z}$ this spiral is liable to flop into a transverse conical spiral,
propagating along $z$ and canted toward $x.$ Furthermore, it was also
shown \cite{ghimire} that spin fluctuations in form of a helical magnon
propagating along $y$ can be selectively excited, generating a topological
field (and hence the topological Hall effect) proportional to the temperature
and also dependent on the net magnetization. In Ref. \cite{ghimire} a simple
formula was derived, which reads
\begin{equation}
{\rho}^{T}=\kappa(1-M^{2}/M_{s}^{2})TH, \label{RT}%
\end{equation}
where $\kappa$ is an unknown, material-specific constant, and $M_{s}$ is the
saturation magnetization.

However, direct substitution of Eq. \ref{RT} into Eq. \ref{RH} is not
possible, for the reason that the assumption that $R_{0}$ and $R_{s}$ do not
depend on magnetic field is, while popular, generally incorrect. Both
coefficients are determined by the electronic structure, which, in turn, is
very sensitive to magnetic order. This problem was discussed in Ref.
\cite{ghimire} where the following protocol was worked out: First, the Hall
resistivity in the non-topological phases below (in terms of the external
field $H)$ or above the topological phase ($H_{1}<H<H_{2})$ are fit separately
to the first two terms in Eq. \ref{RH}. In principle, they should be then
continuously connected to each other across the topological region and the
subtracted from the total $\rho^{H}.$ In Ref. \cite{ghimire}, for the lack of
any justifiable recipe, they were simply connected by the straight line. Now,
since the difference, which we will call ${\rho}^{T},$ is, by construction,
zero at $H_{1}$ and $H_{2},$ they subtracted the linear base $\rho
_{0}=[(H-H_{1}){\rho}^{T}(H_{2})+(H_{2}-H){\rho}^{T}(H_{1})]/(H_{2}-H_{1}),$
where ${\rho}^{T}(H)$ was taken from Eq. \ref{RT}.

\begin{figure}[tbh]
\begin{center}
\includegraphics[width=0.99 \columnwidth]{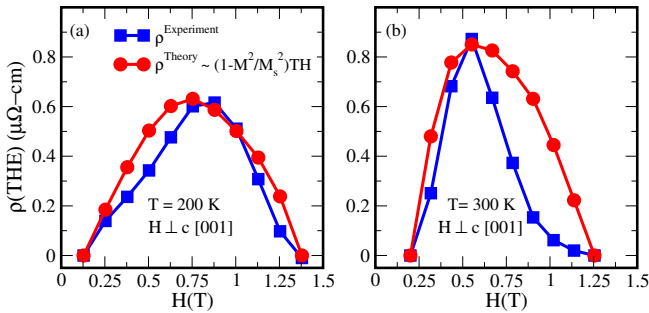}
\end{center}
\caption{Topological Hall effect resisitivity extracted as described in the
text, compared to Eq. \ref{RT}}%
\label{the}%
\end{figure}

We have followed this protocol, albeit the experimental data are not nearly as
clean as in YMn$_{6}$Sn$_{6}$ (Fe$_{3}$Ga$_{4}$ is known to form with
considerable disorder), in particular, proper identification of the first and
the second spin-flop fields is difficult. Still, we were able to tentatively
assign them to be (see Fig. \ref{roab}), at $T=200$ K, $H_{1}\approx0.125$ T
and $H_{2}\approx1.375$ T, and at $T=300$ K, $H_{1}\approx0.18$ T,
$H_{2}\approx1.25$ T (at lower temperatures the topological signal is too weak
to be analyzed quantitatively). The results of this analysis are shown in Fig.
\ref{the}. Note that the amplitude of the topological signal is about 40\%
higher at $T=300$ K, in good agreement with $300/200=1.5$, consistent with the linear dependence on 
$T$ in Eq. \ref{RT}. 

\section{V. Conclusions}

We have studied, using Density Functional Theory, magnetic properties of a
potential topological-Hall material, Fe$_{3}$Ga$_{4}$ metal. We found that the
DFT ground state is a spin spiral, propagating along the crystallographic $c$
direction with $\mathbf{q}=(0,0,0.27)$ reciprocal lattice units. This is in
excellent agreement with the neutron scattering findings for temperatures
above $\sim100$ K. Contrary to the previously published conjecture we
identified this state as a spiral, and not an amplitude spin density wave. We
argue that the actual ground state, despite $b$ being (slightly) the hard
magnetic axis, is an $ab$ helical spiral, stabilized by dipole-dipole
interactions. 

We have further identified a spin flop field at which the helical spiral flops
into a transverse conical spiral, which, according to the theory proposed
recently by one of us for another topological-Hall spiral magnet, YMn$_{6}$Sn$_{6}.$ 
The same theory works well for  Fe$_{3}$Ga$_{4}$. Indeed, the
theory predicts a topological Hall effect in the transverse conical phase only,
with a strong (approximately linear) temperature dependence, and both
predictions are corroborated by the experiment. This, second observation 
of the dynamically fluctuation-induced topological
Hall effect, strongly suggests that the proposed theory is correct and
sufficiently universal.

\section{Acknowledgments}

The authors  acknowledge funding from the U.S. Department of Energy
through Grant No. DE-SC0021089. We also acknowledge the Department of Defense
(DoD) Major Shared Computing Resource Center at Air Force Research Laboratory
(AFRL) and the National Energy Research Scientific Computing Center (NERSC)
for high-performance computing facilities, where some portions of the
calculations were performed.  Last but not least, we acknowledge many insightful discussions
with Huibo Cao and Maxim Mostovoy.

\end{document}